\documentclass[aps,prb,twocolumn,showpacs,superscriptaddress,amsmath,amssymb,tightenlines]{revtex4}
\usepackage{graphicx}% Include figure files
\usepackage{dcolumn}% Align table columns on decimal point
\usepackage{bm}% bold math

\usepackage[usenames,dvipsnames]{xcolor}
\usepackage{dsfont}
\usepackage{amsbsy}

\usepackage{amsfonts}
\usepackage{amsmath}
\usepackage{amssymb}
\usepackage{graphicx}
\usepackage{bbold}
\newcommand{\Eq}[1]{Eq.~(\ref{#1})}
\newcommand{\Fig}[1]{Fig.~\ref{#1}}
\newcommand{\Table}[1]{Table~\ref{#1}}

\usepackage{calrsfs}
\DeclareMathAlphabet{\pazocal}{OMS}{zplm}{m}{n}

\newcommand{\PRB}{\textit{Phys. Rev. B }}

\makeatletter
\def\hlinewd#1{%
  \noalign{\ifnum0=`}\fi\hrule \@height #1 \futurelet
   \reserved@a\@xhline}
\makeatother

\begin{document}
\title{Decoherence of nuclear spins in the ``frozen core'' of an electron spin }
\author{R.~Guichard}
\affiliation{Department of Physics and Astronomy, University College London,
Gower Street, London WC1E 6BT, United Kingdom}
\author{S~.J.~Balian}
\affiliation{Department of Physics and Astronomy, University College London,
Gower Street, London WC1E 6BT, United Kingdom}
\author{G.~Wolfowicz}
\affiliation{ Department of Materials, Oxford University, Oxford OX1 3PH, United Kingdom}
\affiliation{London Centre for Nanotechnology, University College London, London WC1H 0AH, United Kingdom}
\author{P.~A.~Mortemousque}
\affiliation{London Centre for Nanotechnology, University College London, London WC1H 0AH, United Kingdom}

\author{T. S. Monteiro}
\affiliation{Department of Physics and Astronomy, University College London,
Gower Street, London WC1E 6BT, United Kingdom}

\date{\today}

\begin{abstract}

Hybrid qubit systems combining electronic spins with nearby (``proximate'') nuclear spin registers
offer a promising avenue towards quantum information processing, with even multi-spin error correction protocols recently demonstrated in diamond.
However, for the important platform offered by spins of donor atoms in cryogenically-cooled silicon,
decoherence mechanisms of $^{29}$Si proximate nuclear spins are not yet well understood.
The reason is partly because proximate spins lie within a so-called ``frozen core'' region
where the donor electronic hyperfine interaction strongly suppresses nuclear dynamics.
We investigate the decoherence of a central proximate nuclear qubit arising from quantum spin baths outside, as well as inside, the frozen core
around the donor electron.
We consider the effect of a very large nuclear spin bath comprising many ($\gtrsim 10^8$) weakly contributing
pairs outside the frozen core. We also propose that there may be an important contribution from a few
(of order $100$) symmetrically sited nuclear spin pairs (``equivalent pairs''), which were not 
previously considered as their effect is negligible outside the 
frozen core. If equivalent pairs represent a measurable source of decoherence,
 nuclear coherence decays could provide sensitive probes
 of the symmetries of electronic wavefunctions.
For the phosphorus donor system,
we obtain $T_{2n}$ values of order 1 second for both the ``far bath'' and ``equivalent pair'' models,
confirming the suitability of proximate nuclei in silicon as
very long-lived spin qubits.

\end{abstract}

% PACS %TODO add one more
%%%%%%%%

% 03.65.Yz
% Decoherence, quantum mechanics

% 03.67.Lx
% Quantum computation

% 76.60.Lz spin echo


\maketitle
\section{Introduction}
The coherent manipulation of quantum spins in either silicon or diamond represents two
promising approaches to the development of a raft of quantum technologies including not only quantum
computing but also sensing, metrology and magnetometry. In diamond, the remarkable properties of nitrogen vacancy (NV) colour centres for spin-dependent optical read-out and polarization are the cornerstones of a large number of proposed applications at the single spin level.\cite{NV1,NV2,NV3,NV4,NV5} In silicon, shallow donors (mainly group V atoms including phosphorus) provide coupled electron-nuclear spin systems which form the basis for the seminal proposal of Kane for scalable silicon-based quantum computing.\cite{Kane1998} There has been much recent progress in single-spin detection and read-out,\cite{Morello2010,Pla2012,Pla2013,Muhonen2014,Pla2014} complementing  studies on ensembles. \cite{Abe2004,Abe2010,Tyryshkin2012,Steger2012} Strong mixing between the donor electronic and nuclear spins leads to
``sweet-spots'' of enhanced electronic spin coherence, even in natural silicon, first investigated theoretically,
\cite{Mohammady2010,Mohammady2012,Balian2012} and also with experiments.\cite{Wolfowicz2013,Balian2014}

Natural silicon comprises mostly spin-free ${}^{28}$Si isotopes, but $4.67\%$ of sites hold ${}^{29}$Si impurities with $I=1/2$ nuclear spins. Similarly, natural diamond is mostly spin-free ${}^{12}$C, but
% the central NV system interacts with the $I=1/2$ nuclear spins of the
 $1.1\%$ of the atomic sites are occupied by ${}^{13}$C nuclei for which $I=1/2$. In these samples,  ${}^{29}$Si or ${}^{13}$C nuclei represent the major source of decoherence.\cite{DeSousa,Witzel1,Witzel2,Maze2008,Zhao2012} However, interest in these impurities has now moved far beyond their role as a destructive source of decoherence to applications
ranging from sensing of a few nuclear spins,\cite{Si29detect,C13detect}  to the very recent demonstrations of quantum registers combining the central electronic qubit with proximate nuclear spins.\cite{Cappellaro,Waldherr,Taminiau}

A pair of proximate nuclear spins can interact not only via direct dipolar coupling, but also via longer-ranged interactions mediated by the central electronic spin. In either case, the nuclear spins of the pair may `flip-flop' and the resulting magnetic noise provides a well-known source of decoherence for both electronic and nuclear qubits. However, in the case of strong hyperfine coupling between the nuclear impurity pair and the electron spin, the resulting energy detuning on each partner of the pair overwhelms the dipolar coupling, suppressing spin flip-flopping and the associated decoherence within a so-called ``frozen core'' region.

The frozen core \cite{Khutsi,Hahn,George} is a well-established concept in electron spin resonance (ESR) studies but
 is now attracting new interest as a reservoir of protected qubits.\cite{diamondbook2013}
The boundary radius $R_\text{FC}$ of the frozen core is commonly
set as the distance at which hyperfine coupling strengths have decreased to values comparable to the dipolar interactions between neighbouring nuclear spins.\cite{diamondbook2013} Representative values of the latter may be inferred from measured linewidths; for example, the $127$~Hz linewidth of $^{29}$Si in natural silicon
\cite{Hayashi} corresponds to an estimated $R_\text{FC} \approx 80$~\AA\ for the phosphorus donor system (Si:P)
which exemplifies our study.

Hyperfine couplings of proximate ${}^{29}$Si {\em sites} in an ensemble were already resolved spectroscopically in 1969 by continuous wave ESR,\cite{HaleMieher}
and their coherence can now be investigated with, for example, pulsed ENDOR techniques.\cite{ENDOR,Balian2012,Wolfowicz2015} For ${}^{29}$Si nuclear spins
 far outside the frozen core the value of the echo decay time  $T_{2n}\simeq 5$ ms  has been measured.\cite{Si5ms}  Very recently,\cite{Pla2014} the $T_{2n}$ of a proximate {\em single} ${}^{29}$Si spin was measured  -- for the case of an ionized donor -- to be 6.4~ms, close to the measured ensemble/bulk value reported in Ref.~\onlinecite{Si5ms}. However, for the single spin,  $T_{2n}$  in the presence of the
neutral donor (i.e. within the frozen core due to the electron) was not measured. This parameter is 
however of interest to potential future realisations using ${}^{29}$Si nuclear spin registers in combination with electronic qubits, analogous to the recent studies of NV$^-$ centres with ${}^{13}$C nuclear spins.\cite{Cappellaro,Waldherr,Taminiau} 

\begin{figure}[t!]
\begin{center}
\includegraphics[width=2.7in]{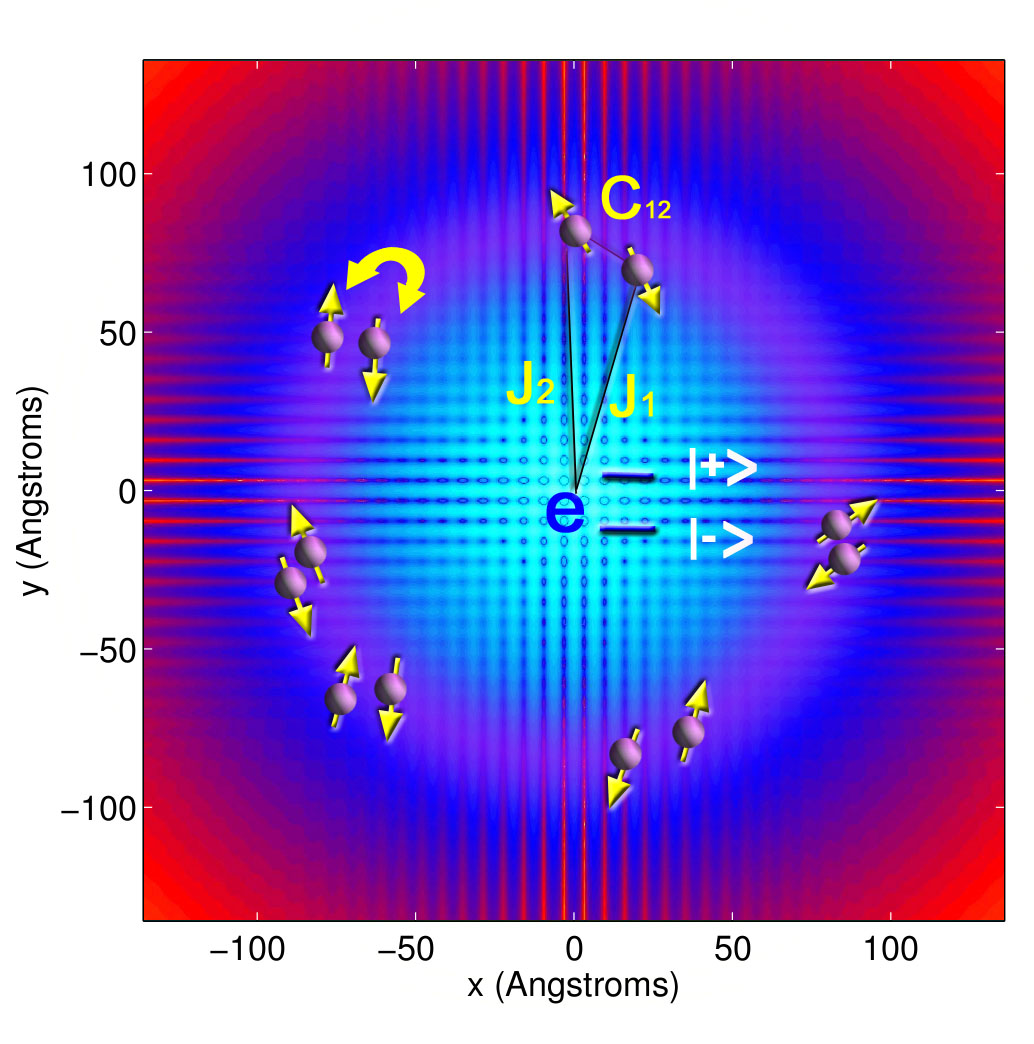}
%\end{center}
\caption{Decoherence of  electronic  spin qubits by a
 flip-flopping nuclear spin bath in natural silicon. The background plots the spatial electronic
 wavefunction (for Si:P here); blue denotes the strong-detuning
  region, where the energy cost of a bath spin flip $\Delta_e^\pm \propto \pm(J_1-J_2)$ exceeds the
strongest intrabath coupling $C_{12}$; it thus corresponds to the usual definition
of the ``frozen-core'' region.  However, electronic spin decoherence is dominated by an active zone (purple colour)
of pairs of  nuclear spins which are actually {\em within} the blue {\em strongly} detuned region,
with  $|\Delta_e^\pm/C_{12}| =|(J_1-J_2)/C_{12}| \sim 10$ for Si:P. 
The reason is that, while for large $|\Delta_e^\pm|$ flip-flop amplitudes are strongly damped,
  qubit state-dependence of the quantum bath evolution, essential for the entanglement between the
electronic spin and bath which produces decoherence,
is also  proportional to $ \Delta_e^\pm$.}
\label{Fig1}
\end{center}
\end{figure}

\begin{figure}[ht!]
%\begin{center}
\includegraphics[width=2.55in]{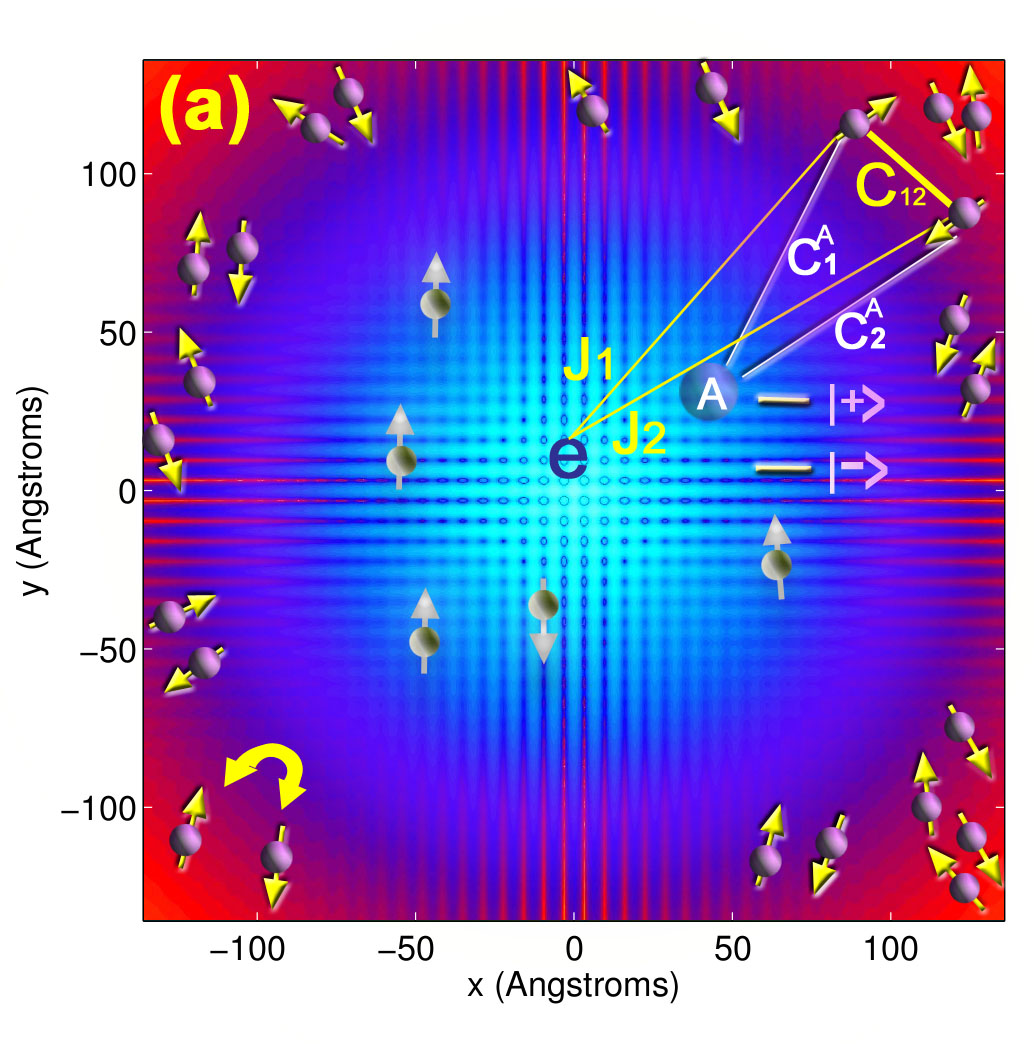}
\includegraphics[width=2.55in]{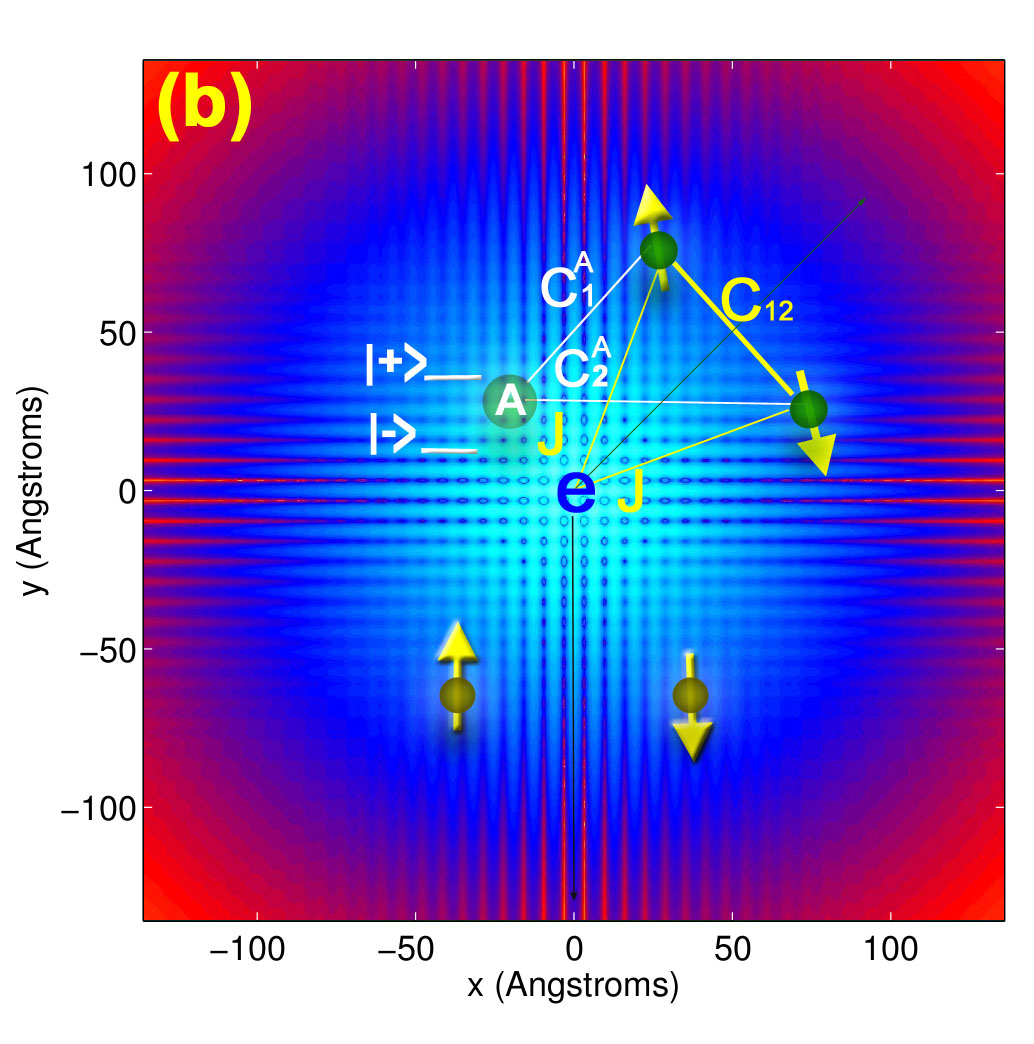}
%\end{center}
\caption{Decoherence of proximate nuclear spins by a quantum bath of nuclear spin pairs  (a) outside and (b) inside the frozen core.
In contrast to electron spin decoherence (for which the detuning is fully state-dependent, see Fig.1), 
the detuning is now $\Delta_e+\Delta_n^{\pm}$: there is  now
 potentially a very large {\em state-independent} component $\Delta_e \propto (J_1-J_2) $ which simply damps the bath noise,
in addition to  a {\em state-dependent} component
 $\Delta_n^{\pm} \propto \pm(C_{1}^A-C_{2}^A)$  which leads to qubit-bath entanglement and thus decoherence.
{\bf (a) Far bath model}  For large $R$, the bath spin interaction with both the electron spin and nuclear 
qubit is dipolar, thus  $|\Delta_n^{\pm}/\Delta_e| \sim 10^{-4}$ so very weak contributions from an extremely large bath of $10^8$ pairs for $R\lesssim 350$~\AA 
~must be combined to obtain a converged decay.
 {\bf (b) Equivalent pairs model}. In the frozen core there are comparatively few spin impurities.
  For equivalent pairs however, $J_1=J_2\equiv J$ so
 $\Delta_e \simeq 0$. Their density is determined by the symmetry of the electronic wavefunction.
The requirement for strong state-selective detuning  implies also that one member of the pair must be
close enough to the qubit to allow appreciable direct dipolar coupling (as opposed to long-range 
coupling between nuclear spins mediated by the electron). Pairs which also satisfy this requirement
 (exemplified by the upper, but not the lower, equivalent pair) are rare but even a few dozen suffice
to exceed the contribution of the $\sim 10^8$ far-bath spin pairs.}
%\end{center}
\label{Fig2}
\end{figure}

Although it has long been known that the large energy detunings in the frozen core drastically suppress nuclear dynamics, 
this regime has not to date been investigated using more recently  developed quantum-bath models, where decoherence is understood  in terms of entanglement between the central spin and bath. This leads to clear differences between decoherence of the electronic and 
proximate qubits, even if in both cases the same nuclear bath is responsible for the decoherence.
 The differences are
summarised in Figs.~\ref{Fig1} and \ref{Fig2} and motivate a more careful look at what is meant by the frozen core and where,
precisely, its boundaries lie. For example, for the electronic qubits, decoherence is in fact dominated \cite{Balian2014} by impurities
which lie within the usual definition of the frozen core, as illustrated in \Fig{Fig1},
 since the detuning fully contributes to entanglement.

In the present study, we consider two decoherence models for the proximate nuclear qubits, either of which, given certain assumptions, might contribute.
Both arise from pair flip-flops of the nuclear impurities, but under quite extreme conditions, not encountered in typical qubit decoherence studies.

{\em (1) Far bath model.}  In this model, we consider the decoherence from distant nuclear spin pairs, which are outside
the frozen core and thus can flip-flop appreciably. We show that the typical contribution is so weak that we must include of order
 $10^8$ flip-flopping pairs outside the frozen core, at distances $R=50-350$~\AA~   from the donor site, in order to obtain results converged with respect to bath size. In contrast, typical quantum-bath calculations of electronic decoherence require  $\sim 10^3- 10^4$ pairs to obtain convergence.

{\em (2) Equivalent pairs model.} In this model, the dephasing noise arises from a few  dozen nuclear spin pairs, well within the frozen core, for which: (i) the members of the pair are symmetrically sited relative to the central spin and thus have equivalent values of the hyperfine detuning (ii) at least one member is sufficiently close to the nuclear qubit to have a significant dipolar interaction, while the other can be remote. They interact via a long-ranged hyperfine interaction mediated by the electron. The indirect flip-flopping of these equivalent pairs (EPs) is found to be most significant,
 but we include also the rarer contribution of direct flip-flops between the nuclear central spin and any equivalent partner it might have. We obtain $T_{2n}$ values in the seconds timescale 
both for individual realisations (relevant to single donor experiments) and also for ensemble averages over many realisations.
Though not previously considered, this  source of dephasing is quite generic: equivalent sites may play a role in any solid state qubit system with a sufficiently dense surrounding nuclear bath.

 Quantum-bath decoherence calculations do not include all combinatorially allowed spin clusters:
there are in principle $\sim 10^{10}$  ${}^{29}$Si spin pairs within $350$~\AA .  Fortunately, only a smaller fraction are physically significant. These are found by numerical search of each randomly populated lattice realization by restricting the selection to, for instance, pairs within a certain distance
and coupling strength. However, applying  normal distance/coupling strength thresholds turned out to be unreliable in the frozen core, so a different strategy had to be adopted to ensure convergence, given the importance of carefully locating the few dozen or so most important EPs which can be quite widely separated.
There is a drastic difference between the choice of spin  clusters which must be included in the quantum bath for each case (a few dozen for the EP model, $\sim 10^8$ for the far bath model); surprisingly, although the decay curves have a different shape, the  $T_{2n}$ values are comparable.

If electronic symmetries are
important, then a strategy for breaking such symmetries, with external fields might be considered to obtain an even longer $T_{2n}$; if the far bath is dominant,  partial isotopic enrichment might be more useful (consideration of ${}^{29}$Si nuclear spin registers in the present study naturally precludes full enrichment).
Either way, the combined
effect is still a $T_{2n} \sim 1$~s even without any other strategy for ameliorating decoherence, such as dynamical decoupling control.
We note that below 5~K, the electronic relaxation time $T_1$ is also above 1~s and would not limit the nuclear spin coherence.

In Section II, we review briefly quantum bath decoherence, for either nuclear or electron qubits.
In Section III.A., we investigate the far bath model and in section III.B we present the EP model.
Finally, we compare $T_{2n}$ values for both models and conclude in Section IV.

\section{Dephasing decoherence of  electronic and nuclear qubits}

The decoherence of the central electronic spin qubit is extremely well-studied.\cite{DeSousa,Maze2008,Renbao2006,Witzel2013} A dephasing process
arising from noise due to flip-flopping nuclear spin pairs is responsible as illustrated in Fig.~1. This basic pair flip-flop mechanism also underlies the 
decoherence of the proximate nuclear spins which are the subject of the present study. Thus it is useful to review briefly the pair decoherence mechanism within the quantum bath approach, for both cases.

The decoherence numerics here employ the CCE (cluster correlation expansion) method for solving for quantum bath decoherence,\cite{Renbao2006} which in general can combine clusters of all sizes, not just pairs. Echo decays of donor electronic spins, in weak electron-nuclear mixing regimes, can be well simulated with 
 just the pair correlation.\cite{Balian2014,Witzel2013} Where the central and bath spins are the same species, higher correlations arising from larger clusters may be required for high accuracy,\cite{Witzel2013} but in both our models would represent only a minor quantitative correction. 
It is only in certain regimes of donors with strong mixing of the central-spin states, \cite{Balian2015} that  a truly qualitative effect 
on the coherence times, arising from larger clusters, is evident.

 For either electronic $(e)$ or nuclear $(n)$ spin decoherence, dephasing arises from entanglement with a quantum bath. The qubit  is prepared in a superposition
of its upper  $\left| + \right\rangle_{n,e}$ and lower $\left| - \right\rangle_{n,e}$  states by an applied resonant $\pi/2$ pulse at $t=0$:
\begin{equation}
\left|\Psi(0)\right>=\tfrac{1}{\sqrt{2}}
\left(\left|+\right>+\left|-\right>\right)_{n,e} \otimes \left| \mathcal{B}(0)\right> \otimes \left| \phi (0)\right\rangle_{e,n}
\label{product}
\end{equation}
where $\left|\phi (0)\right\rangle_{e,n}$ denotes the initial spin state which is {\em not} resonant with the external control pulses, while $\left|\mathcal{B}(0)\right>$ denotes an initial state of the bath. We now consider
 the effect of a single nuclear impurity pair (both $I=1/2$ spins); then, $\left|\mathcal{B}(0)\right>$ is one of the four thermal states: $\{\left|\uparrow\downarrow\right>,\left|\downarrow\uparrow\right>\}$ or $\{\left|\uparrow\uparrow\right>,\left|\downarrow\downarrow\right>\}$. The joint dynamics of the qubit spins and bath are given by:
\begin{equation}
\hat{H}=\hat{H}_{\textrm{q}}+\hat{H}_{\textrm{int}}+\hat{H}_{\textrm{bath}},
\label{HTOT}
\end{equation}
where the qubit Hamiltonian $\hat{H}_{\textrm{q}}= \gamma_e \omega_0\hat{S}_z + \gamma_n \omega_0\hat{I}^{A}_z$ represents the Zeeman terms for the electron and nuclear qubit (labelled $A$), either one of which is resonant with any applied control field. Here we limit ourselves to the unmixed cases where any coupling between the qubit and the host nuclear spin is neglected. The interaction Hamiltonian:
\begin{equation}
 \hat{H}_{\textrm{int}}=\displaystyle\sum_{i=1,2}(\hat{\bf S} \cdot \mathbf{{J}}+ \hat{\bf I}^{A}
 \cdot \mathbf{{C}}^A_{i})
 \cdot \hat{\bf I}_i
\label{Hint}
 \end{equation}
represents the hyperfine coupling $\mathbf{{J}}$ between the central spin and bath spins and the dipolar interaction $\mathbf{{C}}^A_{i}$ between the resonant nuclear qubit and the remainder of the 2-spin bath. For simplicity, we do not include explicitly the term ${\hat{\bf S}} \cdot \mathbf{{J}}\cdot \hat{\bf I}^{A}$ coupling the electron to the resonant nucleus, which is only significant in the specific (but minor) contribution from direct flip-flopping processes.
Finally, the intrabath coupling $\hat{H}_{\textrm{bath}}=\hat{\bf I}_1 \cdot \mathbf{C}_{12} \cdot \hat{\bf I}_2 \simeq
 C_{12}\hat{I}^z_1\hat{I}_2^z-\dfrac{C_{12}}{4}(\hat{I}^+_1\hat{I}_2^-+\hat{I}^-_1\hat{I}_2^+)$, represents the dipolar coupling between
the bath nuclei in secular form (see Appendix A for details).
  
Under the action of the full Hamiltonian \Eq{HTOT} the product state in \Eq{product} evolves into an entangled state:
\begin{equation}
\left|\Psi(t)\right>=
\left[\left|+\right>_{n,e}\left| \mathcal{B}_+(t)\right>+
                                              \left|-\right> _{n,e}  \left| \mathcal{B}_-(t)\right>\right]
                                              \tfrac{\left| \phi(t) \right\rangle_{e,n}} {\sqrt{2}}.
\label{wf}
\end{equation}
The coherence decay is calculated from either $\mathcal{L}_e(t)= \left<\Psi(t)\right| \hat{S}^+\left|\Psi(t)\right>$
or $\mathcal{L}_n(t)= \left<\Psi(t)\right| \hat{I}^+\left|\Psi(t)\right>$. For both cases,
$|\mathcal{L}_{e,n}(t) |\propto |\left<\mathcal{B}_+(t)|\mathcal{B}_-(t)\right>|$,
 so  disregarding a constant prefactor $\left<+\right|\hat{S}^+\left|-\right>$
or $ \left<+\right|\hat{I}^+\left|-\right>$, the coherence has a (real) decaying envelope for the normalized $\ell$-th bath spin pair contribution:
\begin{equation}
\mathcal{L}^{(\ell)}(t) = \left| \left<\mathcal{B}^{(\ell)}_+(t)|\mathcal{B}^{(\ell)}_-(t)\right>\right |  \leq 1.
\end{equation}
The measured  decays, whether Hahn echos or more complex dynamical decoupling sequences,
 are obtained from the combined contributions from many spin pairs:
\begin{equation}
\langle \mathcal{L}(t) \rangle = \langle \prod_\ell  \mathcal{L}^{(\ell)}(t) \rangle
\end{equation}
where the average is over all possible initial bath states, and if required, an ensemble average over different spatial realizations of the  bath. 

An accurate and insightful simplification arises if $\gamma_e \omega_0 \gg |J|$ or if  $\gamma_n \omega_0 \gg |C|$ where $|J|,|C|$ denote typical values of the qubit-bath interaction. In that case, the interaction Hamiltonian in \Eq{Hint} reduces to the Ising form $\hat{H}_{\textrm{int}}=J_1\hat{S}^z\hat{I}_1^z + J_2\hat{S}^z\hat{I}_2^z + C_{1}^A\hat{I}_{A}^z\hat{I}_1^z + C_{2}^A\hat{I}_{A}^z\hat{I}_2^z $, a simple diagonal energy shift in $\hat{I}_1^z$ terms relative to $\hat{I}_2^z$ or a detuning of bath spin 1 relative to bath spin 2. Referred to as the pseudospin model,\cite{Renbao2006} in this case, $\{\left|\uparrow\uparrow\right>,\left|\downarrow\downarrow\right>\}$ bath states do not contribute while the dynamics of the $\{\left|\uparrow\downarrow\right>,\left|\downarrow\uparrow\right>\}$  states is given by an effective state-dependent Hamiltonian:
\begin{equation}
 \hat{H}^\pm= \frac{1}{4}(\Delta^{\pm} \hat{\sigma}_z + C_{12} \hat{\sigma}_x)
\label{equa8}
\end{equation}
with the detuning $\Delta^\pm \simeq\Delta_e^\pm =\pm(J_1-J_2) $ for an electronic qubit and $\Delta^\pm =\Delta_e \pm (C_{1}^A-C_{2}^A)$ for a nuclear qubit. In the latter case, the electronic detuning $\Delta_e \equiv |\Delta_e^\pm|$ represents a potentially large contribution which is not sensitive to the qubit state. From \Eq{equa8} decays (indirect and direct flip-flops) are deduced analytically and for the Hahn echo case we obtain:
\begin{equation}
\mathcal{L}^{(\ell)}(t)\simeq \left|1-2\alpha^{(\ell)}(\alpha^{(\ell)}+i \beta^{(\ell)})\right|,
\label{Decay}
\end{equation}
where $\alpha= \sin (\omega^{+} t)\sin (\omega^{-} t) \sin {(\theta^+ - \theta^-)}$, $\beta=\sin(\omega^+ t) \cos (\omega^- t) \sin \theta^+ + \sin (\omega^- t) \cos (\omega^{+} t) \sin \theta^-$ while  $\theta^\pm=\tan^{-1}\left(C_{12}/\Delta^\pm\right)$ and  the eigenvalues $\omega^\pm = \dfrac{1}{4}\sqrt{(\Delta^\pm )^2+(C_{12})^2}$, and we have dropped the $\ell$ indices for convenience. The larger $\theta^\pm$, the larger the  amplitudes of the flip-flopping of nuclear spin-pairs which drives the decoherence.

When considering the contribution of flip-flopping pairs which are within the frozen core (the EP model below), we obtained excellent
 agreement between the pseudospin equations above and full numerical CCE  provided that the well-known perturbative correction for the  non-Ising (anisotropic) hyperfine terms,\cite{Renbao2006} was added to the dipolar coupling when using \Eq{Decay}. In other words, for the $\ell$-th cluster,
in Equations~(\ref{equa8}) and (\ref{Decay}) we take:
\begin{equation}
C_{12} = C^{(D)}_{12} + \frac{J_{1} J_{2}}{\omega_0}
\label{mediate}
\end{equation}
\noindent where, $C^{(D)}_{12}$ is the dipolar interaction between nuclear spins, 
allowing bath spins a long-ranged interaction mediated by the central electron, the second term in \Eq{mediate}.
 For a full CCE calculation, $C_{12} = C^{(D)}_{12} $ since the additional effective long-ranged interaction
emerges naturally if the full hyperfine interaction is included.
For non-equivalent pairs in the frozen core, $|\Delta^\pm| \simeq \Delta_e \gg |C_{12}|$, 
thus $\theta^\pm\simeq 0$ and flip-flops become too strongly suppressed.  
The qubit state sensitivity enters in \Eq{Decay} mainly
 through the $\sin {(\theta^+ - \theta^-)}$ 
prefactor and is also suppressed by $\Delta_e$.
 This imposes the further condition  $|\Delta^\pm_n|=|(C_{1}^A-C_{2}^A)| \gtrsim \Delta_e$ for a single individual pair to contribute apppreciably to the decay.

Central to our modelling is the identification of spin clusters within the frozen core which can contribute non-negligibly to the decoherence of a proximate spin.
We now consider two models and  apply them to the particular case of natural Si:P.

%**********************************************
\section{Decoherence of proximate spin qubits}
\subsection{\textbf {Far bath model}}

From numerical simulations with a very large spin bath, we find that
distant spin pairs outside the frozen core radius $R_{\textrm{FC}}$ individually make an extremely small
contribution to decoherence: the $\alpha \propto \sin {(\theta^+ - \theta^-)}$ prefactor
 scales the coherence decays in \Eq{Decay}, since $ |\mathcal{L}| \sim 1- \alpha^2(..)$.
We can also show that the approximate weight of the $\ell$-th pair, is of order
$(1 / T_2^{(\ell)} )^2 \propto  \sin^2 {(\theta_{(\ell)}^+ - \theta_{(\ell)}^-)},$\cite{Balian2014}
assuming also the temporal character of the associated magnetic noise is relevant: in other words,  
flip-flop frequencies $\omega$ for the given pair cannot be orders of magnitude different from $ \sim 1/T_2$.
For a non-negligible contribution we would expect that $N_p |\sin {(\theta^+ - \theta^-)}|^2 \sim 1$
where $N_p$ is a representative number of contributing spin pairs.

 From the pseudospin model (leaving out the $\ell$ labels),
$\sin {(\theta^+ - \theta^-)} \simeq \frac{2C_{12}}{\omega}  \ \frac{C_{1}^A-C_{2}^A}{\omega}$, 
since $\omega^\pm \simeq \omega =\frac{1}{4}\sqrt{\Delta_e^2+C_{12}^2}$.
The first factor ($\frac{2C_{12}}{\omega}$) determines whether the pair can flip-flop
appreciably and is significant if $|C_{12}/ \Delta_{e}| \sim 1$. The second factor, $\frac{(C_{1}^A-C_{2}^A)}{\omega}$ determines state
distinguishability.
For the far spins, the hyperfine mediated correction plays little role
since $J_1$ and $J_2$ are small. 
For distances $\gtrsim 100$~\AA,    where the Fermi contact component of the hyperfine interaction becomes small, the 
dipolar electronic-nuclear hyperfine interaction still makes a contribution to the detuning which is much
larger than $(C_{1}^A-C_{2}^A)$. Here,

\begin{equation}
\frac{(C_{1}^A-C_{2}^A)}{\omega} \sim \frac{(C_{1}^A-C_{2}^A)}{J_1-J_2} \sim 10^{-4} \simeq \frac{\gamma_n}{\gamma_e}
\end{equation}
Thus the contribution of each such far bath spin pair  is
 $\left(\frac{\gamma_n}{\gamma_e}\right )^2 \sim 10^{-8}$, so
 only a far bath with $N_p \sim 10^{8}$ contributing spin pairs can produce significant decay.  At very large $R$, however,
$(C_{1}^A-C_{2}^A)/\omega  \to (C_{1}^A-C_{2}^A)/C_{12}$ 
But there is a minimum value  of the interaction $C_{12}\equiv C^{min}_{12}$ where $(C^{min}_{12})^{-1}$  sets a  timescale
below which the bath noise is too slow to contribute. As $R \to \infty$, then $(C_{1}^A-C_{2}^A)/C^{min}_{12} \to 0$
thus there is a maximum radius $R_{max}$ beyond which the far bath does not contribute significantly to decoherence.

\begin{figure}[t!]
\begin{center}
\includegraphics[width=3.3in]{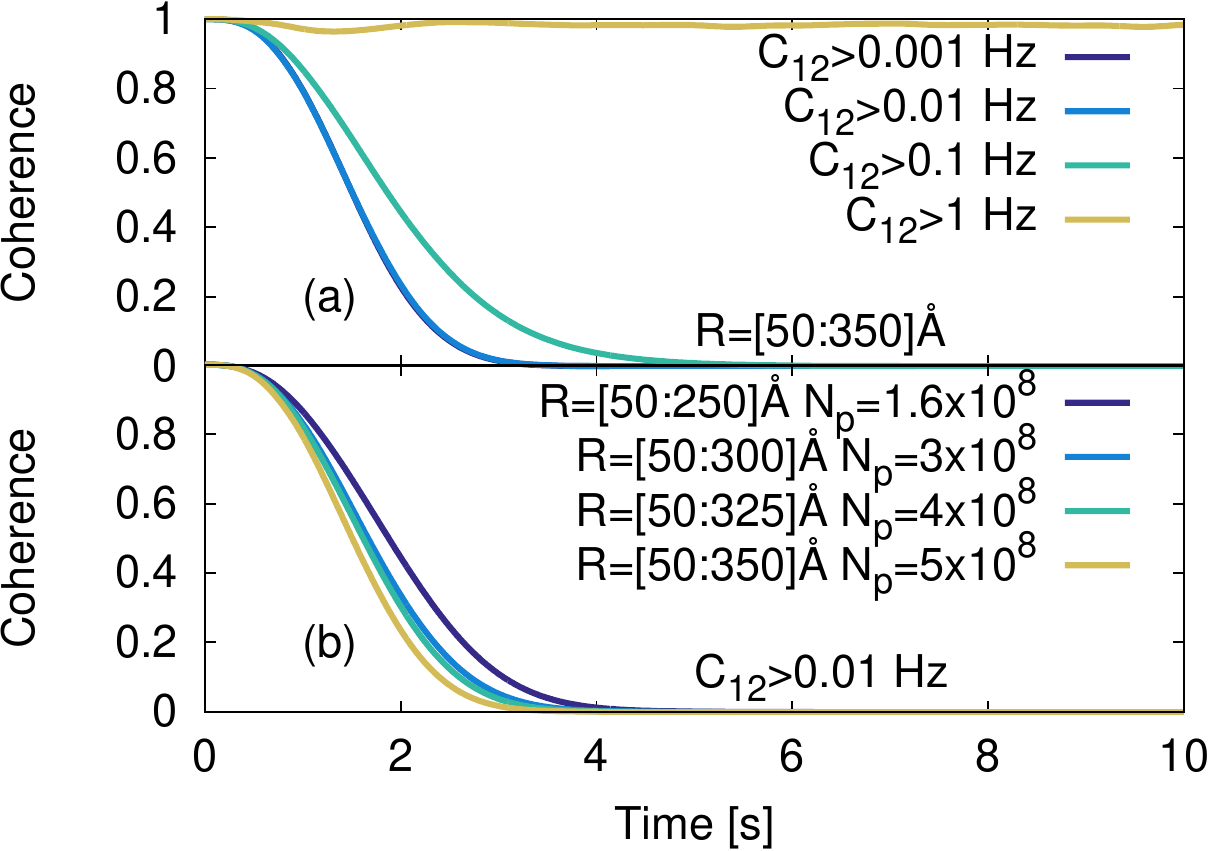}
\end{center}
\caption{Convergence of large bath model with respect to intrabath dipolar coupling (a) and  with respect 
to bath size (b). The figure indicates that decoherence is dominated by spins with $C_{12} \sim 0.01 -1$~Hz and
a bath of spins within $R \lesssim 350$~\AA~ of the origin, combining the contributions from $5 \times 10^8$ spin pairs.
Calculations were performed for the case of Si:P, for $X$-band and magnetic field orientation $B_0=[1 0 0]$, yielding a $T_{2n}$ of 2 s for a
single nuclear $^{29}$Si spin sited at the origin. This represents an estimate for the upper bound for the coherence time
if the far bath is the dominant process.
Due to the large nuclear spin bath, the coherence decays are insensitive to the choice of random spatial realization of the bath.
}
\label{Fig3}
\end{figure}

We tested this analysis numerically by means of CCE calculations using a very large bath of nuclear spin pairs
(excluding EPs) and testing the effect on coherence decays of increasing the size of the bath.
Figure~\ref{Fig3} shows convergence with respect to bath size for Hahn echo decays, for a nuclear spin at the origin (thus
expected to give an upper bound on the coherence).  The $C_{12}\equiv C^{min}_{12} \sim 0.01-0.1$ Hz bound indicates that the pairs are within
$40-50$~\AA~ of each other and the calculation  is converged with respect to bath size if we include $5 \times 10^8$ spin pairs within $R_{max}  \lesssim 350$~\AA~ of the origin.  
The scale of the bath is remarkable, in comparison with comparable electronic decoherence calculations with $\sim 10^4$ pairs.
%Equally remarkable is that, in the converged bath, each bath impurity can participate in potentially 
% $\sim 1000$ flip-flopping pairs with spins at other sites.

Although it is computationally
feasible to solve for a bath of this magnitude by CCE2 or pseudospin methods,  the uniformity of the bath means that
it is reliable to evaluate $\mathcal{L}$ in a smaller but geometrically representative sample of the bath. In addition
no averaging over bath realisations was required: the results are insensitive to whether one has a single spin or an ensemble.

% (say $F$) and then to evaluate $(\mathcal{L})^F$

 \subsection{ \textbf{Equivalent pairs model}} 

The isotropic part of the hyperfine interaction is modeled using the Kohn-Luttinger (KL) wavefunction which is an approximation of the ground state wavefunction of the donor electron developed onto Bloch wavefunctions and parametrized by the experimental ionization energy; for phosphorus donors, $E_{\textrm{p}} = 0.044$ eV (see Appendix A for details).
We can estimate the local densities of suitable EPs in the isotropic case before considering effects from any anisotropies in the hyperfine coupling.

\begin{figure}[ht!]
\begin{center}
\includegraphics[width=0.9\linewidth]{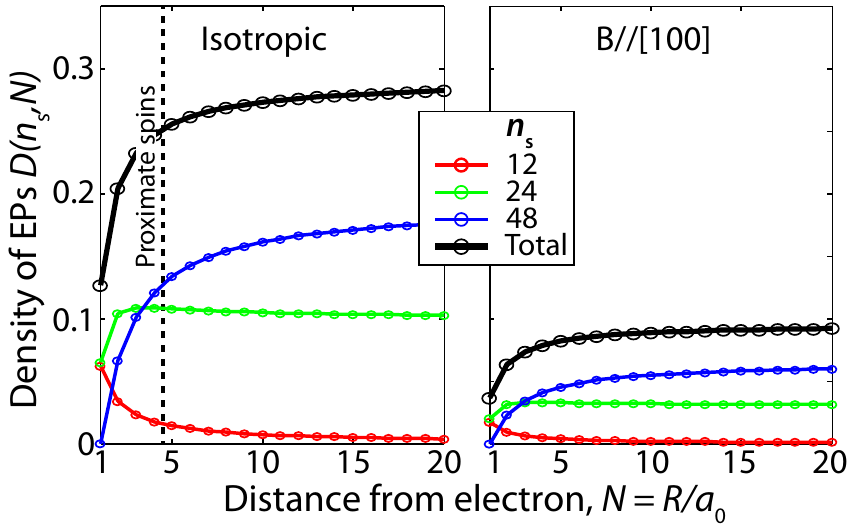}
\end{center}
\caption{Density of equivalent pairs (EPs) as a function of distance. The separate contributions from different types of shells is shown as well as the total density assuming a purely isotropic contact interaction (left) or a correction for anisotropic behaviour (right). The density of EPs is approximately constant for $ R \gtrsim 10$~\AA,
but the innermost proximate spins typically interact with fewer EPs.}
\label{fig4}
\end{figure}

The Si crystal structure can be described by a simple cubic lattice with a parameter $a_0=5.43$~\AA\ and an 8-site basis. All atomic sites are represented by an integer vector ${\bf n} = [n_1,n_2,n_3]$. In our simulations, the full lattice size ranges over $[-N,N]$ cubic cells for each dimension, resulting in 8$N^3$ unit cells and hence 64$N^3$ total atomic sites. Owing to the symmetry of the system, each site possesses several potential \textit{equivalent} partners, for which positions can be deduced from any allowed permutations of $[\pm n_1,\pm n_2,\pm n_3]$. We can assign each vector $ \bf{n}$ to a shell $s$ comprising $n_s=48,24,12,8,6$ or $4$ partners and we first obtain $\mathcal{N}_{n_s}(N)$, the number of shells comprising $n_s$ partners within a radius of $R=Na_0$ from the center. The ranges were adjusted to ensure summation over {\em complete} shells (see Appendix B for details) and we obtain estimates for $\mathcal{N}_{n_s}(N)$:
\begin{eqnarray}
\mathcal{N}_{12}(N)&=&4N^2\nonumber\\
\mathcal{N}_{24}(N)&=&\frac{4}{3}N(N^2-1)+N^2\nonumber\\
\mathcal{N}_{48}(N)&=& \frac{2}{3} N^3-N^2+\frac{N}{3},
\label{sitenumb}
\end{eqnarray}
\noindent while $\mathcal{N}_{8}(N)=\mathcal{N}_{6}(N)=N, \mathcal{N}_{4}(N)=2N$. Then assuming a binomial distribution, taking an abundance of $p=0.0467$ for nuclear spin impurities in natural silicon, the estimated average number of significant EP in each shell is:
\begin{eqnarray}
\overline{\zeta}_{n_s} \simeq \sum_k \binom{n_s}{k} p^k (1-p)^{n_s-k} \ \frac{k(k-1)}{2},
\label{avpair}
\end{eqnarray}
\noindent where $ \binom{n}{k}=\frac{n!}{k!(n-k)!} $ is the binomial coefficient. For the two dominant shells $\overline{\zeta}_{48} \sim 2.3$ and $\overline{\zeta}_{24} \simeq 0.6$. In these cases, it is quite likely that any impurity spin has an equivalent partner {\em somewhere}, albeit remotely located. Nevertheless, due to the long-range electron-mediated coupling, a $C_{12}$ of about tens of Hz is present. Within a sphere of radius $N$ cubic cells, we expect the total number of EP to be simply:
 \begin{eqnarray}
 \mathcal{N}_{\textrm{EP}} \simeq \sum_{s} \overline{\zeta}_{n_s} \mathcal{N}_{n_s}(N).
\label{avpair2}
\end{eqnarray}
\noindent For instance, within a radius of $R=100$~\AA,  we find $\mathcal{N}_{\textrm{EP}} \simeq 19,000$. For a proximate nucleus however, one member of the pair must be dipolar-coupled to the resonant spin (cf. caption of \Fig{Fig1}) which is relevant within about $m\sim 3$ cubic cells. Each nuclear qubit thus interacts with other nuclei in the neighbouring $(2m)^3 \sim 200$ cells. We can define a density of spin pairs:
\begin{eqnarray}
D(n_s,N=R/a_0)= \frac{\overline{\zeta}_{n_s}\mathcal{N}_{n_s}(N)}{(2N)^3},
\label{loc}
\end{eqnarray}
\noindent which gives the mean number of EPs in each cubic cell as a function of distance, $R=Na_0$ from the electron. We see in \Fig{fig4} that the mean number for large $R$ is about $0.2-0.3$ pairs per cubic cell (cf. left panel), thus each nuclear qubit interacts with $\sim 50$ potential EPs if anisotropy is neglected.

\begin{figure}[t!]
\begin{center}
\includegraphics[width=3.3in]{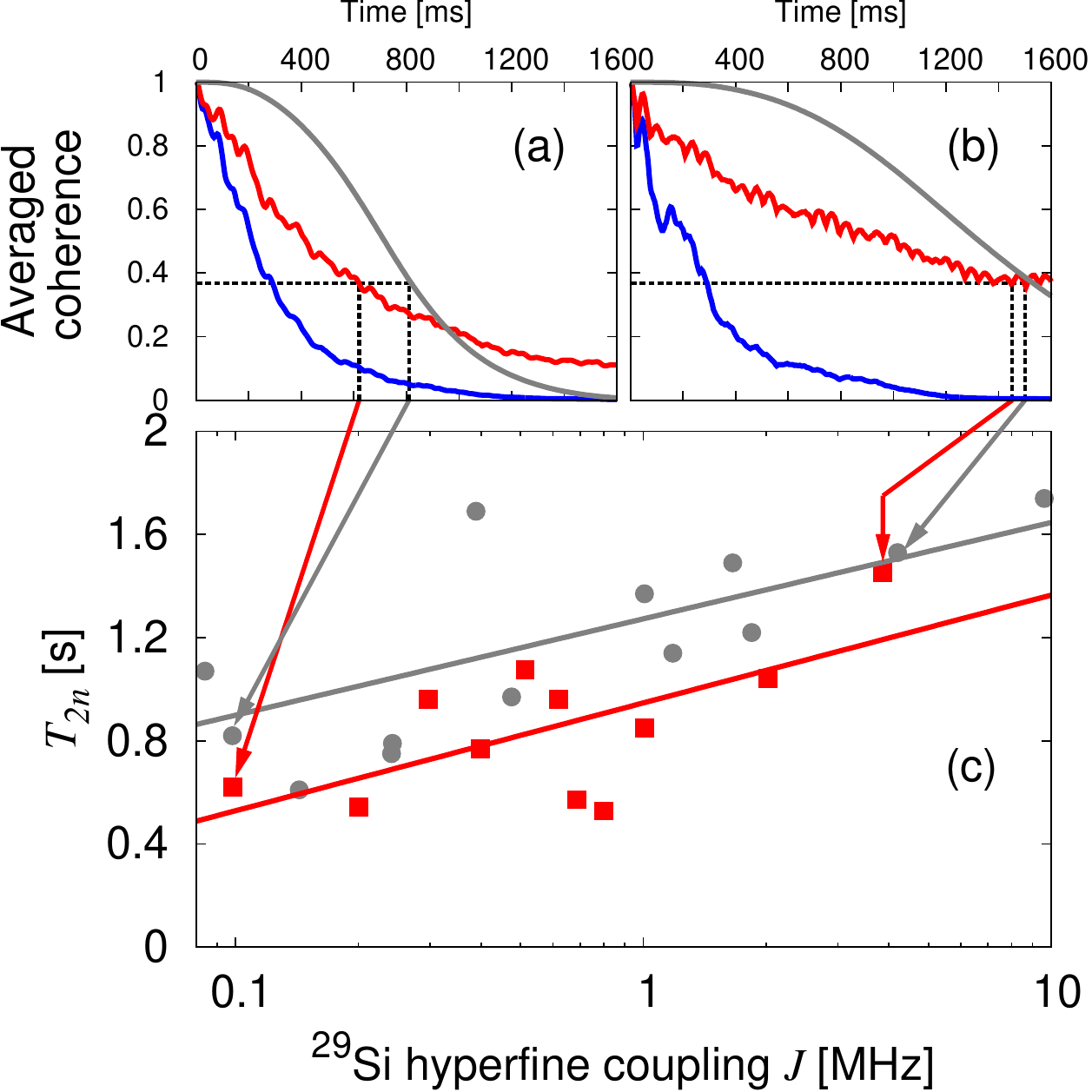}
\end{center}
\caption{(Colour online) Top panels:  calculated Hahn echo decays for proximate spins 
 in a \textsuperscript{nat}Si:P system for  {\bf (a)} $J = 0.1$ MHz and {\bf(b)} $J = 3.8$ MHz;   blue: isotropic hyperfine coupling; red: includes anisotropy correction.  {\bf (c)} Calculated $T_{2n}$ values.
There is a weak trend for $T_{2n}$ to increase as the hyperfine coupling increases (red line is a fit), possibly indicative of the decreasing density of EPs as $R \to 0$. In the far bath model (grey dots), the slight increase in decoherence with lower $J$ (grey line) reflects the fact that the lower $J$ proximate spins are slightly
closer to the far bath.
Coherence times were obtained from decays averaged over 100 spatial realizations of the bath, but typical single realizations gave the same timescale of decoherence.
}
\label{Fig5}
\end{figure}

For the numerical calculations of the echo decays, we carried out a careful search, retaining about 500 equivalent spin pairs and
 averaging over 100 realisations of a randomly generated lattice population with $4.67\%$ 
of sites occupied by ${}^{29}$Si spins. Two sets of calculations of the Hahn echo decays were carried out.
The first employed only the isotropic contact interaction and neglected anisotropic components of the hyperfine interaction. These calculations provide a lower bound for the  $T_{2n}$ and predicted  decay rates $T_{2n} \sim 0.2-0.3$~s for different values of $J$ (see Appendix C for further details).

A second set of calculations attempts to account for the anisotropy, which is less easy to calculate reliably. We assumed that any degree of anisotropy detunes spin pairs so much that their contribution became negligible. In effect, this model provides an upper bound for the expected  $T_{2n}$ as not all shells are affected by anisotropy.
Thus to remain an equivalent pair we required that spins have the same $(\hat{\bf{n}}_B \cdot \bf{n})^2$, where $\hat{\bf{n}}_B$ is the direction of the magnetic field. The effect is to reduce the symmetries but to increase the number of shells, i.e.\ for $\hat{\bf{n}}_B = [1,0,0]$ and the main shells with $n_s=48, 24, 12$ partners, we have $n_s \to n_s/3$ and therefore $\mathcal{N}_{n_s} \to 3\mathcal{N}_{n_s}$ (cf. \Fig{fig4} right panel). 

\section{Coherence decays}

We have calculated coherence decays for a representative set of proximate nuclear spins by comparing
the effect of a large far bath model with a new model introduced here, based on the effect of
symmetrically-sited spin pairs, the equivalent pairs (EPs).
The corresponding $T_{2n}$ are shown in \Fig{Fig5} along with the comparable far bath results.
 We  found that, in both cases, $T_{2n}$  is of order 1 second
with a weak  dependence on $J$ -- in both cases, the coherence times tend to increase with larger $J$,
a trend also seen in recent experiments.\cite{Wolfowicz2015}
 
For the EP model we treat the anisotropic
correction simply as a symmetry lowering effect; this is plausible as the resultant detuning
would be extremely large. Presently, it is not possible  to fully include anisotropy using the KL wavefunction; typically, the dipolar
correction within this framework is included with a Heaviside function,\cite{DeSousa} and is thus neglected for $R \lesssim 20$\AA. Given other uncertainties,
these two EP calculations provide an upper and a lower bound to $T_{2n}$. As both results are on the seconds timescales, they suffice
for the practical aim of establishing the proximate nuclear spins as useful qubits. 

If the Si:P wavefunction exhibits a degree of spatial 
symmetry comparable with the KL wavefunction,
then the EPs could be the dominant mechanism, albeit only slightly.
However, it is likely that such symmetries are at least partly broken; in that case, the far 
bath would limit $T_{2n}$. Given the uncertainties in the KL wavefunction, at present
it is not possible to determine accurately the contributions of EPs relative to the far bath,
but as -- fortuitously -- the timescales are comparable, one can still conclude that the resulting $T_{2n}$ is about 1 second.

To facilitate comparison with ensemble experiments, the EP results are averaged over many realizations
(the far bath model coherence decays are fully insensitive to ensemble averaging).
In the EP model, decoherence is primarily due to an indirect flip-flop process and this arises from several dozen such EPs.
Thus, although results from single donors  fluctuate between realisations, the corresponding order of magnitude for $T_2$
 remains on the one second timescale, whether ensemble averaging is carried out or not. The exception is the atypical realization
where the central spin happens to have an equivalent site it can directly flip-flop with. For proximate central spins, usually
 in inner shells with
 $n_s=4,8,12$ this is unlikely. We find that the small subset of such realizations decohere rapidly. They contribute little to
the ensemble averaged $T_{2n} \sim 1$~s values but would clearly be unsuitable as qubit registers unless some strategy to exploit the degeneracy is envisaged. 

\section{Conclusion}

We have investigated the coherence of nuclear spins lying within the so-called ``frozen core''
within a quantum bath framework.   We have calculated the coherence using a very large far bath of spins lying
outside the frozen core.  
We also introduced here a new model, based on equivalent pairs (EPs) deep within the so-called ``frozen core'', which we argue
would  limit the phase coherence of proximate nuclear spins -- provided the electronic wavefunctions has the symmetries of the KL wavefunction (or even an alternative model with  comparable levels of symmetry).
 Within the EP model, decoherence is primarily due to an indirect flip-flop process arising from a few dozen such EPs.
Our quantitative results are  indicative: a more refined investigation should consider improved wavefunctions, more accurate than the KL wavefunction.\cite{Pica} Experimental investigation 
of the behaviour including dependence on symmetry-breaking mechanisms (such as crystal orientation and strain) will be useful to test this proposal.

Our models predicts $^{29}$Si nuclear spin coherence times of order 1 second (using only a Hahn echo),
 which is consistent with recent experimental measurements,\cite{Wolfowicz2015} showing such spins could be useful as potential qubits.

{\em Acknowledgements}
We are indebted to Kohei Itoh, Steve Lyon, Brendon Lovett, John Morton, Giuseppe Pica, Jarryd Pla  and Martin Uhrin for useful discussions. R.~ G. is supported by EPSRC Collaborative Computational Project Q grant EP/J010529/1. S.~J.~B. is supported by the Stocklin-Selmoni Studentship through the UCL Impact Programme.

\appendix

\section{Dipolar and hyperfine interactions}

The dipolar interaction strength $C_{ij}$ between two nuclear spins $i$ and $j$
with gyromagnetic ratios $\gamma_i$ and $\gamma_j$, defined in the main text ($C_{12}$) is given by
\begin{equation}
C_{ij} (r_{ij},\gamma_i,\gamma_j,\theta_{ij}) =
\frac{\mu_0}{4\pi} \gamma_i \gamma_j \hbar \frac{(1 - 3 \cos^2{\theta_{ij}})}{r_{ij}^3},
\end{equation}
where $r_{ij}$ the inter-nuclear separation,
$\mu_0 = 4\pi \times 10^{-7}$~NA$^{-2}$ and $\theta_{ij}$ denotes the angle between
the magnetic field and the line connecting the two nuclei.\cite{DeSousa}

The hyperfine interaction is the magnetic interaction between an electronic spin
$\hat{\bf S}$ and a localized nucleus $\hat{\bf I}$.
Due to the large mismatch between electronic and 
nuclear gyromagnetic ratios and a sufficiently
strong magnetic field, it can be written in secular form:\cite{DeSousa}
\begin{equation}
\hat{H}_{\text{HF}} = \left(J({\bf r}) - C_{en}(r,\gamma_e,\gamma_n,\theta) \Theta(r - r_0) \right) \hat{S}^z \hat{I}^z,
\label{fullhyperfine}
\end{equation}
where ${\bf r}$ is the electron-nuclear separation vector, $r \equiv |{\bf r}|$
and the electron-nuclear dipolar interaction $C_{en}$ is non-zero for $r > r_0$ ($r_0 \approx 20$~\AA\ for Si:P).

The first term in \Eq{fullhyperfine} is the {\em isotropic} Fermi contact interaction.
For a donor electron interacting with a nucleus, the strength $J$
is given by (see e.g.\ Ref.~\onlinecite{DeSousa})
\begin{align}
J({\bf r})=P & [ F_{1}({\bf r})\cos{(k_{0}x)}  \nonumber \\
&+ F_{2}({\bf r})\cos{(k_{0}y)} + F_{3}({\bf r})\cos{(k_{0}z)}]^{2}.
\end{align}
Here, ${\bf r} = (x,y,z)$ and $k_0 = (0.85)2\pi/a_0$
with lattice constant $a_0$.
The prefactor $P$ contains the electronic and nuclear
gyromagnetic ratios ($\gamma_{e}$ and $\gamma_{n}$),
and the charge density on each atomic site $\eta$,
\begin{equation}
P=\frac{4}{9}\gamma_{S} \gamma_{I}\hbar\eta\mu_0.
\end{equation}
The relevant envelope functions of the Kohn-Luttinger (KL) model wavefunction are:
\begin{equation}
F_{1}({\mathbf r}) = \frac{\exp{\left[
-\sqrt{\frac{x^{2}}{(nb)^{2}}+\frac{y^{2}+z^{2}}{(na)^{2}}}\right]}}
{\sqrt{\pi(na)^{2}(nb)}},
\end{equation}
\begin{equation}
F_{2}({\mathbf r}) = F_{1}({\mathbf r}) \text{~with~} \{x \to y, y \to z, z \to x\},
\end{equation}
\begin{equation}
F_{3}({\mathbf r}) = F_{1}({\mathbf r}) \text{~with~} \{x \to z, y \to x, z \to y\},
\end{equation}
where $a$ and $b$ are lengths characteristic to the donor
and $n = \sqrt{0.029~\text{eV}/E_i}$ with the electron ionization energy $E_i$ in eV.

\begin{table}[b]
\begin{center}
  \begin{tabular}{ c c c c }
    \hlinewd{1pt}
     $n_s$ & 48 & 24 & 12 \\
    \hline
    Class&&& \\
    1 & $24N^2(N-1)$ & $12N(3N-1)$ & $12N$ \\
     &&& \\
    2 & $8N(N-1)(N-2)$ & $36N(N-1)$ & $12N$ \\
    &&& \\
    3 & -- & $16N(N-1)(2N-1)$ & $24N(2N-1)$\\
    \hlinewd{1pt}
  \end{tabular}
  \caption{Class contribution to the equivalent sites group as a function of $N$.}
  \label{tab1}
\end{center}
\end{table}

\begin{figure}[t!]
\begin{center}
\includegraphics[width=3.375in]{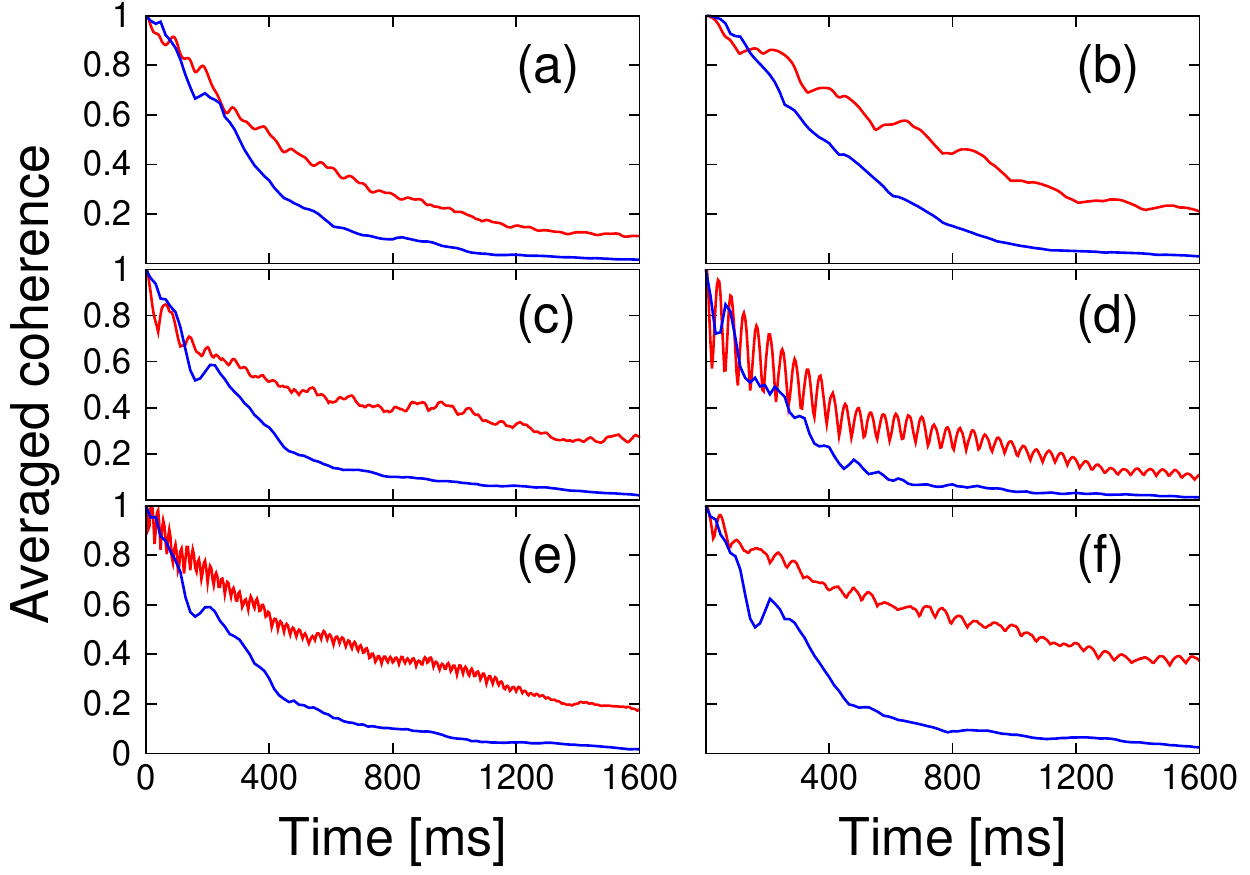}
\end{center}
\caption{Simulations of decays the coherence decays of a set of proximate spins
corresponding to $J = 0.1$ (a),
$J = 0.3$ (b), $J = 0.5$ (c), $J = 0.7$ (d), $J = 1$ (e), $J = 3.8$ (f) MHz.
The blue lines correspond to isotropic coupling only and yield $T_{2n}\approx 0.2-0.3$ s;
red lines show the effect of symmetry reduction
due to the anisotropy of couplings: we compare the effect of the further desymmetrisation
if we constrain EP to have in addition the same
orientation condition (same $(\hat{\bf{n}}_B \cdot \bf{n})^2$ as discussed in main text).
The effect is to produce $T_{2n}$ in the seconds timescale.}
\label{Fig6}
\end{figure}

\section{Counting equivalent sites}

Equivalent sites are those with the same hyperfine interaction which we obtain using the KL model
wavefunction described above.
We begin by considering the allowed coordinates of the impurities in the crystal.
The Si crystal structure can be described by a simple cubic structure with lattice parameter $a_0=5.43$~\AA\ and an 8-site basis.
All atomic sites are represented by an integer vector ${\bf n} = [n_1,n_2,n_3]$ which are obtained from translations (modulo 4) in all directions of the 8 basis vectors [0,0,0], [0,2,2], [2,0,2], [2,2,0], [3,3,3], [3,1,1], [1,3,1], [1,1,3].
For simplicity, we can sort these vectors into three classes: class 1 contains [0,2,2], [2,0,2], [2,2,0], class 2 contains [0,0,0] and class 3 [3,3,3], [3,1,1], [1,3,1], [1,1,3]. To ensure counting over complete shells described by the basis vectors, summations must range between $[-N,N]$ for the 2 coordinates and $[-N,N-1]$ for the 0 coordinate of class 1 giving $4N^2(2N+1)$ number of sites; between $[-N,N]$ for class 2 giving $(2N+1)^3$ number of sites and between $[-N,N-1]$ for class 3 giving $8N^3$ number of sites. Owing to the symmetry of the system, each site possesses several \textit{equivalent} partners with positions which can be deduced by permutations of $[n_1,n_2,n_3]$ and which lie on the surface of shells of radius $R=\frac{a_0}{4}\sqrt{n_1^2+n_2^2+n_3^2}$. By consideration of the symmetries of the KL wavefunction we can assign each vector $ \bf{n}$ to a group of $n_s=48,24,12,8,6$ or $4$ partners. For each class, the contribution to a shell comprising $n_s$ partners within a radius of $R=Na_0$ of the center as a function of $N$ is summarized in \Table{tab1}.
Additionally, class 2 contributes as $8N$ to $n_s=8$ and as $6N$ to $n_s=6$ and class 3 contributes as $8N$ to $n_s=4$. 

%\section{Desymmetrisation due to anisotropy}

Finally, in \Fig{Fig6} we compare the effect of the further desymmetrisation
if we constrain EP to have in addition the same
anisotropy correction, i.e.\ same $(\hat{\bf{n}}_B \cdot {\bf n})^2$.

\end{document}